\documentstyle[graphics,graphpap]{article}
\begin{document}
\title{Detection statistics in the micromaser}
\author{David B. Johnson and W.C. Schieve}
\date{\today}
\maketitle


\begin{abstract}
We present a general method for the derivation of various statistical quantities describing the detection of a beam
of atoms emerging from a micromaser.  The use of non-normalized conditioned density operators and a linear master
equation for the dynamics between detection events is discussed as are the counting statistics, sequence statistics,
and waiting time statistics.  In particular, we derive expressions for the mean number of successive detections of
atoms in one of any two orthogonal states of the two-level atom.  We also derive expressions for the mean waiting times
between detections.  We show that the mean waiting times between detections of atoms in like states are equivalent to
the mean waiting times calculated from the uncorrelated steady state detection rates, though like atoms are indeed
correlated.  The mean waiting times between detections of atoms in unlike states exhibit correlations.  We evaluate the
expressions for various detector efficiencies using numerical integration, reporting results for the standard
micromaser arrangement in which the cavity is pumped by excited atoms and the excitation levels of the emerging atoms
are measured.  In addition, the atomic inversion and the Fano-Mandel function for the detection of de-excited atoms is
calculated for comparison to the experimental results of Weidinger~{\it et~al.}~\cite{Weidinger1999}, which reports the
first observation of trapping states.
\end{abstract}

\section{Introduction}

The most fundamental system for studying matter-radiation coupling is a single two-level atom interacting with a
single mode of an electromagnetic field---first treated theoretically by Jaynes and Cummings \cite{Jaynes1963}.  The
one-atom-maser or micromaser~\cite{Meschede1985} is an experimental realization of this system.  Other applications
for the Jaynes-Cummings model include ion traps~\cite{Cirac1996}, quantum nondemolition
measurements~\cite{Brune1990,Holland1991,Brune1992}, quantum state teleportation~\cite{Davidovich1994,Cirac1994},
quantum computation~\cite{Sleator1995}, and optical cavity QED~\cite{Frezberger1999}.  For a nice review of the early
experiments, see Raithel {\it et~al.}~\cite{Raithel1994}.  For more recent discussions,
consult~\cite{Haroche1997,Scully1997,Englert1998}.


Much work has been done on determining just how much information about the micromaser cavity field can be obtained from
the detection statistics of the emerging atoms.  In the standard micromaser experiments, atoms enter the high-$Q$
resonator in the upper of the two Rydberg states of the maser transition and the emerging atoms are probed for being in
either of these two states.  For this case, the detection statistics of the emerging atoms are sensitive only to the
photon-number distribution in steady state.  The detection statistics have been shown to contain information on the
photon statistics~\cite{Rempe1990} and the intensity fluctuations~\cite{Herzog1994,Cresser1994,Herzog1995,Cresser1996}. 
In order to measure field coherence properties, means for breaking the phase symmetry of the field have been
proposed---both by coherently pumping the cavity~\cite{Krause1986,Scully1991,McGowan1997} and by a measurement induced
breaking of the phase symmetry~\cite{Wagner1992,Wagner1992b,Wagner1993}.  Both arrangements can be
used to measure the phase diffusion rate and linewidth of the field, but neither has been shown to give the full
details of the spectrum.  Cresser~\cite{Cresser2000} has recently shown that all of the information needed to obtain
the full spectrum is contained in the atomic beam and has proposed a Mach-Zehnder type interferometer for measuring the
general micromaser spectrum.

Most of the efforts to make connections between the detection statistics and the field properties has concentrated on
the correlation functions for the detection of de-excited atoms, which are not conditioned to exclude detections between
events.  Knowledge of the whole set of correlation functions would provide a complete description of the statistical
properties from which all statistical quantities of interest could be calculated in principle; however, many
statistical quantities (e.g.\ waiting times) are much easier to derive from the conditioned evolution of the cavity
field between detections.

The conditioned evolution of the cavity field between detections must include the effect of the passage of atoms
undetected due to detector inefficiency.  As shown by Briegel {\it et~al.}~\cite{Briegel1994}, this leads to a
nonlinear master equation for the conditioned evolution of the field between detections.  However, a linear master
equation and a non-normalized conditioned density operator, as used by Herzog~\cite{Herzog1994}, can be used in place of
the nonlinear equation and the normalized conditioned density operator.  In Sec.~\ref{detStats} of this paper, we will
show how the two methods are equivalent and review the methods of calculating detection statistics in the micromaser
in general.

Using a perturbation method based on the nonlinear equation of Ref.~\cite{Briegel1994}, McGowan and Schieve have derived
an effective micromaser spectrum as calculated from the known detection statistics---one that agrees with the true
spectrum in the case of perfect detector efficiency, but otherwise includes a back-action effect due to measurement. 
Herzog~\cite{Herzog2000} points out that this is not the conventional micromaser spectrum; usually defined as the
Fourier transform of the steady state normalized two-time correlation function of the electric field strength.  The
differences between the two definitions for the spectrum is an interesting question for the future that will not be
discussed in this paper.

In Sec.~\ref{observables}, we discuss three types of statistical quantities: counting statistics, sequence statistics,
and waiting time statistics.  In Sec.~\ref{count} we will discuss counting statistics and, in particular, the
Fano-Mandel function for measuring the difference in the variance of the counts from that of Poissonian statistics. 
They were treated rather nicely by Briegel {\it et~al.}~\cite{Briegel1994} and will be discussed here only briefly.

In Sec.~\ref{sequence}, we will discuss sequence statistics.  Sequence statistics are fundamentally characterized by
the probabilities for a given number of successive detection events being of atoms in a particular sequence of the
states $|{\rm A}\rangle$ and $|{\rm B}\rangle$, where $|{\rm A}\rangle$ and $|{\rm B}\rangle$ are any two orthogonal
states of the two-level atom.  In Sec.~\ref{succsdets}, they will be used to derive expressions for the mean number of
successive detections of atoms in state $|{\rm A}\rangle$ and the mean number of successive detections of atoms in
state $|{\rm B}\rangle$.  The method used is analogous to the one used by Englert {\it et~al.}~\cite{Englert1996} to
derive an expression for the mean number of successive detections of atoms in the same state.

Waiting time statistics will be discussed in Sec.~\ref{wait}.  There we will derive expressions for the mean waiting
times between successive detections.  We show that the mean waiting times between detections of atoms in the same state
are equivalent to those calculated from the uncorrelated steady state detection rates.  Of course atoms in the same
state are correlated, and we show that the simplification that leads to the uncorrelated mean waiting times between
atoms in the same state does not occur for mean squared waiting times and presumably for higher powers of waiting
times between atoms in the same state.  We also derive expressions for the mean waiting times between atoms in unlike
states, which do exhibit correlations.

In Sec.~\ref{numerical}, we will discuss a numerical integration method based on the linear master equation which can
be used to evaluate the expressions derived in Sec.~\ref{observables}, and in Sec.~\ref{standSetup} we report results
for the standard micromaser arrangement.  In addition to evaluating the expressions derived in Sec.~\ref{observables},
we calculate the atomic inversion for comparison to the new experimental results of Weidinger {\it
et~al.}~\cite{Weidinger1999} who recently reported the first observation of trapping states.  We find that there is
qualitative agreement.  In an attempt to improve the agreement, we included an uncertainty in the interaction time with
a Gaussian distribution, among other effects, but failed to improve the agreement significantly.  We also evaluate the
Fano-Mandel function for the detection of de-excited atoms for comparison to the results of Weidinger {\it et~al.}  We
find poor agreement due in part to the level of amplitude suppression exhibited by the experimental results, presumably
due to some effect not included in our calculation.  We evaluate the mean number of successive detections of atoms in
each state and the mean waiting times between detections of atoms in unlike states, and find that they are strongly
affected by the presence of trapping states---offering an alternative means of the experimental observation of trapping
states.

\section{Micromaser Dynamics}

The density operator of the micromaser cavity field evolves as
\begin{eqnarray}
\rho(t)=e^{{\cal X}rt}\,\rho(0)\ ,
\label{Evolution}
\end{eqnarray}
where $r$ is the rate of atomic injection with assumed Poissonian statistics, and the evolution operator ${\cal X}$ is
given by
\begin{eqnarray}
{\cal X}={\cal L}+{\cal M}(\varphi)-1\ .
\end{eqnarray}
Here ${\cal L}$ is the Liouvillian operator that governs thermal damping of the field, and ${\cal M}(\varphi)$ is a
super operator that governs the change in the field due to the passage of a single two-level atom with accumulated Rabi
angle $\varphi$.

The field damping operator ${\cal L}$ is given by~\cite{Scully1997}
\begin{eqnarray}
{\cal L}\rho=-{1\over2N_{\rm ex}}\left[(\nu+1)\left(a^\dagger a\rho-2a\rho a^\dagger+\rho a^\dagger
a\right)+\nu\left(aa^\dagger\rho-2a^\dagger\rho a+\rho aa^\dagger\right)\right]\ ,
\end{eqnarray}
where $N_{\rm ex}\equiv r/\gamma$ for convenience, $\nu$ is the mean number of thermal photons, and $\gamma$ is the
photon decay rate.  Note that for any $\rho$, ${\rm tr}\{{\cal L}\rho\}=0$.

The atomic passage operator ${\cal M}(\varphi)$ results from solving the atom-field interaction problem for the
combined atom-field density operator $\rho_{\rm a-f}$ and then tracing over the atomic subsystem.  By performing the
trace in a basis spanned by two orthogonal states $|{\rm A}\rangle$ and $|{\rm B}\rangle$ of the two-level atom, $\cal
M$ can be written as ${\cal M}={\cal A}+{\cal B}$ where ${\cal A}\rho=\langle{\rm A}|\rho_{\rm a-f}|{\rm A}\rangle$ and
${\cal B}\rho=\langle{\rm B}|\rho_{\rm a-f}|{\rm B}\rangle$.  The states $|{\rm A}\rangle$ and $|{\rm B}\rangle$ may be
the Rydberg states of the maser transition, or any coherent superposition of those states.  In this paper we will
consider only the case in which all of the atoms arrive in the upper maser level.  Then ${\cal M}$ is given by
\cite{Scully1997}
\begin{eqnarray}
{\cal M}\rho=\cos\left(\varphi\sqrt{aa^\dagger}\right)\,\rho\,\cos\left(\varphi\sqrt{aa^\dagger}\right)
+\hat{\gamma}^\dagger\,\sin\left(\varphi\sqrt{aa^\dagger}\right)\,\rho\,\sin\left(\varphi\sqrt{aa^\dagger}\right)\,\hat{\gamma}\ , \label{m}
\end{eqnarray} with the normalized raising and lowering operators
\begin{eqnarray}
\hat{\gamma}^\dagger&\equiv&a^\dagger{1\over\sqrt{aa^\dagger}}={1\over\sqrt{a^\dagger a}}a^\dagger\ ,\nonumber\\
\hat{\gamma}&\equiv&a{1\over\sqrt{a^\dagger a}}={1\over{\sqrt{aa^\dagger}}}a\ .
\end{eqnarray}
The form of ${\cal M}$ is different for the general case (in which the atoms arrive in a coherent superposition). 
Note that for any $\rho$, ${\rm tr}\{{\cal M}\rho\}={\rm tr}\{\rho\}$.

The steady state density operator is defined by ${\cal X}\rho^{\rm ss}=0$.  When all of the atoms arrive in the upper
maser level (${\cal M}$ given by Eq.~(\ref{m})), $\rho^{\rm ss}$ is given by Ref.~\cite{Filipowicz1986}
\begin{eqnarray}
\rho^{\rm ss}(a^\dagger a)=\rho^{\rm ss}(0)\,\prod_{j=1}^{a^\dagger
a}\left[{\nu\over\nu+1}+{N_{\rm ex}\over\nu+1}{\sin^2(\varphi\sqrt{j})\over j}\right]\ , \label{steady}
\end{eqnarray}
where $\rho^{\rm ss}(0)$ is determined by the normalization condition ${\rm tr}\{\rho^{\rm ss}\}=1$.

The accumulated Rabi angle is given by $\varphi=gt_{\rm int}$, where $t_{\rm int}$ is the interaction-time and
\begin{eqnarray}
g={1\over L}\int_0^Ldx\,g(x)\ ,
\end{eqnarray}
is the atom-field coupling constant given in terms of the position dependent atom-field coupling strength $g(x)$ for a
cavity of length $L$.

As described by Meystre {\it et~al.}~\cite{Filipowicz1986,Meystre1988}, in a cavity field with $n_0$ photons, each atom
will undergo an integer number $q$ of Rabi cycles if $\varphi$ obeys the equation
\begin{eqnarray}
\varphi(n_0,q)={q\pi\over\sqrt{n_0+1}}\ .
\end{eqnarray}
When this equation is satisfied, each atom leaves the cavity photon number unchanged, and the photon number is
``trapped'' with $n_0$ being the maximum number of cavity photons allowed (except for thermal noise).  These trapping
states will be shown to strongly affect the statistical quantities derived in Sec.~\ref{observables}.

\section{Detection Statistics}
\label{detStats}

For a particular detection event of interest, it is convenient to define the operator ${\cal X}^+$ corresponding to
the occurrence of that detection event, and the operator ${\cal X}^-\equiv{\cal X}-{\cal X}^+$ corresponding to
evolution of the cavity field in the absence of detection events of that type.  For example, if we are interested in
the detection of outgoing atoms in state $|{\rm A}\rangle$, then ${\cal X}^+=\eta_{\rm A}{\cal A}$ and ${\cal
X}^-={\cal L}+(1-\eta_{\rm A}){\cal A}+{\cal B}-1$, where $\eta_{\rm A}$ denotes the detector efficiency.

A detection event at time $t$ leads to the non-local state reduction
\begin{eqnarray}
\rho(t)\rightarrow {{\cal X}^+\rho(t)\over {\rm tr}\{{\cal X}^+\rho(t)\}}\ ,
\end{eqnarray}
and occurs with a probability density
\begin{eqnarray}
r\,{\rm tr}\{{\cal X}^+\,\rho(t)\}\ .
\end{eqnarray}

For times between detections, the passing of an atom at time $t$, undetected due to detector inefficiency, leads to the
non-local state reduction
\begin{eqnarray}
\rho(t)\rightarrow {({\cal M}-{\cal X}^+)\rho(t)\over{\rm tr}\{({\cal M}-{\cal X}^+)\rho(t)\}}\ ,
\end{eqnarray}
and occurs with a probability density
\begin{eqnarray}
r\,{\rm tr}\{({\cal M}-{\cal X}^+)\,\rho(t)\}\ .
\end{eqnarray}
This, along with the continuous effect of thermal damping, leads to the master equation for the evolution of the
density operator between detections conditioned by the absence of detections
\begin{eqnarray}
{1\over r}\dot{\rho_{\rm c}}&=&{\cal L}\rho_{\rm c}+{\rm tr}\{({\cal M}-{\cal X}^+)\rho_{\rm c}\}\left({({\cal M}-{\cal
X}^+)\rho_{\rm c}\over{\rm tr}\{({\cal M}-{\cal X}^+\rho_{\rm c}\}}-\rho_{\rm c}\right)\nonumber\\
&=&{\cal X}^-\rho_{\rm c}-{\rm tr}\{{\cal X}^-\rho_{\rm c}\}\,\rho_{\rm c}\ ,
\label{conditionedMaster}
\end{eqnarray}
where we have used the identities ${\rm tr}\{{\cal M}\rho_{\rm c}\}=1$, ${\cal X}^+={\cal X}-{\cal X}^-$, and, for any
$\rho$, ${\rm tr}\{{\cal X}\rho\}=0$.  This was derived by Briegel {\it et~al.}~\cite{Briegel1994} for the case ${\cal
X}^+=\eta_{\rm A}{\cal A}+\eta_{\rm B}{\cal B}$, corresponding to the detection of atoms in either state (Eq.~(2.25) of
that paper).  This equation is nonlinear, and has the formal solution given by Eq.~(2.26) of Ref.~\cite{Briegel1994}
\begin{eqnarray}
\rho_{\rm c}(t) = {e^{{\cal X}^-rt}\,\rho(0)\over {\rm tr}\{e^{{\cal X}^-rt}\,\rho(0)\}}\ .
\label{conditionedEvolution}
\end{eqnarray}
The nonlinearity in Eq.~(\ref{conditionedMaster}) is what leads to the normalization of the conditioned density
operator (\ref{conditionedEvolution}).

The conditional probability density for a detection occurring at time $t$ given that no detections have occurred
between times $0$ and $t$ is given by
\begin{eqnarray}
r\,{\rm tr}\{{\cal X}^+\rho_{\rm c}(t)\}\ .
\label{conditionedDensity}
\end{eqnarray}

The {\em exclusion probability}, or the probability that a detection does not occur between times $0$ and $t$ is
\begin{eqnarray}
\exp\left(-r\int_0^tdt\,{\rm tr}\{{\cal X}^+\rho_{\rm c}(t)\}\right)=\exp\left(-r\int_0^tdt\,{{\rm tr}\{{\cal
X}^+e^{{\cal X}^-rt}\,\rho(0)\}\over{\rm tr}\{e^{{\cal X}^-rt}\,\rho(0)\}}\right)\ .
\label{exclusionProbabilityStart}
\end{eqnarray}
If we define
\begin{eqnarray}
u(t)\equiv{\rm tr}\{e^{{\cal X}^-rt}\,\rho(0)\}\ ,
\label{exclusionProbability}
\end{eqnarray}
We find that
\begin{eqnarray}
\dot{u}(t)=r\,{\rm tr}\{{\cal X}^-e^{{\cal X}^-rt}\,\rho(0)\}=-r\,{\rm tr}\{{\cal X}^+e^{{\cal X}^-rt}\,\rho(0)\}\ ,
\end{eqnarray}
where we have used ${\cal X}^-={\cal X}-{\cal X}^+$ and, for any $\rho$, ${\rm tr}\{{\cal X}\rho\}=0$.  So the
expression~(\ref{exclusionProbabilityStart}) becomes
\begin{eqnarray}
\exp\left(\int\limits_0^tdt\,{\dot{u}(t)\over u(t)}\right)=u(t)\ .
\end{eqnarray}
Thus the exclusion probability is $u(t)$ given by Eq.~(\ref{exclusionProbability}) and is equivalent to the denominator
of Eq.~(\ref{conditionedEvolution}), as was noted by Englert {\it et~al.}~\cite{Englert1996}.

The probability density for the next detection occurring at time $t$ is the conditional probability
density~(\ref{conditionedDensity}) for a detection occurring at time $t$ multiplied by the probability $u(t)$ that the
condition is satisfied, or
\begin{eqnarray}
u(t)\,r\,{\rm tr}\{{\cal X}^+\rho_{\rm c}(t)\}=r\,{\rm tr}\{{\cal X}^+{\tilde{\rho}}_{\rm c}(t)\}\ .
\label{next}
\end{eqnarray}
Here we have defined the non-normalized conditioned density operator
\begin{eqnarray}
{\tilde{\rho}}_{\rm c}(t)\equiv u(t)\rho_{\rm c}(t)=e^{{\cal X}^-rt}\,\rho(0)\ ,
\label{noNormConditionedEvolution}
\end{eqnarray}
where the decay in the norm of the operator is due solely to the exclusion of detections between times $0$ and $t$. 
This is the solution to the linear master equation
\begin{eqnarray}
{1\over r}{\dot{\tilde{\rho}}}_{\rm c}={\cal X}^-{\tilde{\rho}}_{\rm c}\ .
\label{noNormConditionedMaster}
\end{eqnarray}
This has been used by Herzog~\cite{Herzog1994}.  The two approaches to Eq.~(\ref{next}) are identical, but the linear
equation (\ref{noNormConditionedMaster}), as opposed to the nonlinear equation (\ref{conditionedMaster}), is simpler to
numerically integrate (see Sec.~\ref{numerical} for a discussion of our numerical integration method).

In the following sections, the subscripts ``A'', ``B'', and ``AB'' will denote detections of atoms in state $|{\rm
A}\rangle$, state $|{\rm B}\rangle$, and state $|{\rm A}\rangle$ or $|{\rm B}\rangle$ (e.g. ${\cal X}_{\rm AB}^+\equiv
\eta_{\rm A}{\cal A}+\eta_{\rm B}{\cal B}$).

\section{Observables of Interest}
\label{observables}

Here we will discuss three types of statistical quantities: counting statistics, sequence statistics, and
waiting time statistics.  Counting statistics and the Fano-Mandel function were treated rather nicely by Briegel {\it
et~al.}~\cite{Briegel1994} and will be discussed in Sec.~\ref{count} only briefly.  Sequence statistics will be
discussed in Sec.~\ref{sequence} and, In Sec.~\ref{succsdets}, will be used to derive expressions for the mean number
of successive detections.  In Sec.~\ref{wait}, we will derive the mean waiting times between successive detections. 
In Sec.~\ref{numerical}, we will discuss a numerical integration method which can be used to evaluate the expressions. 

Ignoring prior observations, detections of atoms in state $|{\rm A}\rangle$ and $|{\rm B}\rangle$ will occur at the
uncorrelated rates
\begin{eqnarray}
r_{\rm A}=r\,{\rm tr}\{{\cal X}_{\rm A}^+\rho^{\rm ss}\}\ ,\ \ r_{\rm B}=r\,{\rm tr}\{{\cal X}_{\rm B}^+\rho^{\rm ss}\}\ .
\label{rates}
\end{eqnarray}
For any statistical variable $x$, it is interesting to calculate the average normalized to the uncorrelated value
\begin{eqnarray}
\langle x\rangle_{\rm norm}\equiv{\langle x\rangle\over\langle x\rangle_{\rm uncor}}\ ,
\end{eqnarray}
where $\langle x\rangle$ is the average of $x$, and $\langle x\rangle_{\rm uncor}$ is the average performed in the
absence of correlations with the detection rates (\ref{rates}).  This ratio is equal to unity for Poissonian
statistics, and otherwise tells us whether the correlations are positive or negative, and, for the case of successive
detections, whether the detection events are bunched or antibunched.

As the detectors become less efficient, the times between detections grow larger (with an increasing number of atoms
passing undetected between detections).  The detected atoms thus become less correlated and we have the general result
\begin{eqnarray}
	{\rm for\ any\ }x\ ,\ \ \langle x\rangle_{\rm norm}\rightarrow1\ ,\ \ {\rm as\ }\eta_{\rm A}, \eta_{\rm
B}\rightarrow0\ .
\label{lessCorelations}
\end{eqnarray}

\subsection{Counting statistics}
\label{count}

The counting statistics were discussed by Briegel {\it et~al.}~\cite{Briegel1994}.  By counting statistics, we mean
the statical properties of the detection events counted for a certain observation period.  They are fundamentally
characterized by the joint probabilities for detecting $n$ atoms in state $|{\rm A}\rangle$ and $m$ atoms in state
$|{\rm B}\rangle$ in an observation time $t$.  Of particular interest is the Fano-Mandel function $Q$ which measures the
deviation of the variance of the counts from that for Poissonian statistics~\cite{Fano1947,Mandel1979}.  It is defined
by convention as \begin{eqnarray} Q(t)&\equiv&\left\langle(N(t)-\langle N(t)\rangle)^2\right\rangle_{\rm norm}-1\
,\nonumber\\ &=&{\left\langle N(t)^2\right\rangle\over\langle N(t)\rangle}-\langle N(t)\rangle-1\ , \end{eqnarray}
where $N(t)$ is the number of counts in an observation time $t$.  An approximate expression for calculating the
Fano-Mandel function was developed by Rempe and Walther~\cite{Rempe1990}, who achieved results that agreed well with a
computer simulation, and later with experimental results~\cite{Rempe1990b}.  An exact expression was derived by
Briegel~\cite{Briegel1994}, and was shown to agree well with the previous approximation in the regime for which it is
valid.  For counting atoms in state $|{\rm B}\rangle$ in an observation time $t$, the Fano-Mandel function can be
expressed as~\cite{Briegel1994} \begin{eqnarray} Q_{\rm B}(t)=&&2r\int_0^tdt'\left(1-{t'\over t}\right)\,{\rm
tr}\left\{{\cal X}_{\rm B}^+e^{{\cal X}rt'}{\cal M}_{\rm B}\rho^{\rm ss}\right\}\ , \label{Qb}
\end{eqnarray}
where
\begin{eqnarray}
{\cal M}_{\rm B}\rho\equiv{{\cal X}_{\rm B}^+\rho\over{\rm tr}\{{\cal X}_{\rm B}^+\rho\}}-\rho\ .
\label{Mb}
\end{eqnarray}
The Fano-Mandel function $Q_{\rm A}$ for the detection of atoms in state $|{\rm A}\rangle$ can be obtained by replacing
${\cal X}_{\rm B}^+$ with ${\cal X}_{\rm A}^+$ in Eqs.~(\ref{Qb}) and~(\ref{Mb}).  $Q_{\rm B}$ is related to the
two-time correlation function for the detection of atoms in state $|{\rm B}\rangle$~\cite{Briegel1994}.  The
Fano-Mandel function $Q_{\rm field}$ for the number of photons in the cavity can be related to the short-time limit of
the same atomic correlation function~\cite{Englert1998}, providing a connection between $Q_{\rm B}$ and $Q_{\rm
field}$.

\subsection{Sequence statistics}
\label{sequence}

Here we ask a different question: ``given $n$ successive detections, what is the probability that the detections are of
atoms in a particular sequence of the states $|{\rm A}\rangle$ and $|{\rm B}\rangle$?''  We use the notation $P[{\rm
ABBA}\ldots(n\ {\rm terms})]$ to denote the probability of an $n$-detection event being comprised of atoms detected in
the sequence of states ``${\rm ABBA}\ldots.$''  They are normalized such that, for each $n$, the sum of all $P[x]$,
where $x$ is a sequence of ``{\rm A}'' and ``{\rm B}'' of length $n$, is unity.  The simplest of these probabilities
are the probabilities that a single atom is detected in state $|{\rm A}\rangle$ or $|{\rm B}\rangle$, given by
\begin{eqnarray}
P[{\rm A}]={{\rm tr}\{{\cal X}_{\rm A}^+\rho^{\rm ss}\}\over{\rm tr}\{{\cal X}_{\rm AB}^+\rho^{\rm ss}\}}\ ,\ \
P[{\rm B}]={{\rm tr}\{{\cal X}_{\rm B}^+\rho^{\rm ss}\}\over{\rm tr}\{{\cal X}_{\rm AB}^+\rho^{\rm ss}\}}\ .
\end{eqnarray}

At the next level of complexity are the probabilities $P[{\rm AA}]$, $P[{\rm BB}]$, $P[{\rm AB}]$, and $P[{\rm
BA}]$ for $n=2$.  Rather than calculate them directly, it is convenient to first consider the conditional probabilities
$P[\underline{\rm A}{\rm A}]$, $P[\underline{\rm B}{\rm B}]$, $P[\underline{\rm A}{\rm B}]$, and $P[\underline{\rm
B}{\rm A}]$, where the underline indicates the given condition.  For example, $P[\underline{\rm A}{\rm B}]$ denotes
the probability that the second detection of a two-detection event is of an atom in state $|{\rm B}\rangle$, given
that the first detection is of an atom in state $|{\rm A}\rangle$.  The joint probability $P[{\rm AB}]$, for example,
is then given by the conditional probability $P[\underline{\rm A}{\rm B}]$ multiplied by the probability $P[{\rm A}]$
that the condition is satisfied:
\begin{eqnarray}
P[{\rm AB}]=P[{\rm A}]\,P[\underline{\rm A}{\rm B}]\ .
\end{eqnarray}
Note that this relationship between joint probability and conditional probability is true regardless of what condition
is made.  For example, it is also true that $P[{\rm AB}]=P[{\rm B}]\,P[{\rm A}\underline{\rm B}]$.

We will now derive an expression for $P[\underline{\rm A}{\rm A}]$.  Expressions for the other conditional
probabilities can be generated in an analogous manner.  Given that an atom in state $|{\rm A}\rangle$ was detected at
time $0$ we have $\rho(0)={\cal X}_{\rm A}^+\rho^{\rm ss}/{\rm tr}\{{\cal X}_{\rm A}^+\rho^{\rm ss}\}$.  Until a later
time $t$, when a second atom in state $|{\rm A}\rangle$ is detected, we whish to exclude detections of atoms in either
state, so we use the non-normalized conditioned density operator ${\tilde{\rho}}_{\rm c}(t)=e^{{\cal X}_{\rm
AB}^-rt}\rho(0)$.  Then the probability density for the time $t$ when the second atom is detected is $r\,{\rm
tr}\{{\cal X}_{\rm A}^+{\tilde{\rho}}_{\rm c}(t)\}$, so that
\begin{eqnarray}
P[\underline{\rm A}{\rm A}]=r\int_0^\infty dt\,{\rm tr}\{{\cal X}_{\rm A}^+{\tilde{\rho}}_{\rm c}(t)\}={1\over{\rm
tr}\{{\cal X}_{\rm A}^+\rho^{\rm ss}\}}{\rm tr}\left\{{\cal X}_{\rm A}^+{-1\over{\cal X}_{\rm AB}^-}{\cal X}_{\rm
A}^+\rho^{\rm ss}\right\}\ .
\end{eqnarray}
The identities
\begin{eqnarray}
&&{\cal X}\rho^{\rm ss}=0\ ,\nonumber\\
&&{\rm for\ any\ }\rho\ ,\ \ {\rm tr}\{{\cal X}\rho\}=0\ ,
\label{identities}
\end{eqnarray}
are especially useful.  Using ${\cal X}_{\rm A}^+={\cal X}-{\cal X}_{\rm AB}^--{\cal X}_{\rm B}^+$ and the identities
(\ref{identities}), we have
\begin{eqnarray}
{\rm tr}\left\{{\cal X}_{\rm A}^+{-1\over{\cal X}_{\rm AB}^-}{\cal X}_{\rm A}^+\rho^{\rm ss}\right\}={\rm tr}\{{\cal
X}_{\rm A}^+\}-\Gamma
\end{eqnarray}
and
\begin{eqnarray}
P[\underline{\rm A}{\rm A}]=1-{\Gamma\over{\rm tr}\{{\cal X}_{\rm A}^+\rho^{\rm ss}\}}\ ,
\end{eqnarray}
where we have defined
\begin{eqnarray}
\Gamma\equiv{\rm tr}\left\{{\cal X}_{\rm A}^+{-1\over{\cal X}_{\rm AB}^-}{\cal X}_{\rm B}^+\rho^{\rm ss}\right\}
={\rm tr}\left\{{\cal X}_{\rm B}^+{-1\over{\cal X}_{\rm AB}^-}{\cal X}_{\rm A}^+\rho^{\rm ss}\right\}\ .
\end{eqnarray}

The sequence probability P[{\rm AA}] and the other sequence probabilities for $n=2$ are given by
\begin{eqnarray}
&&P[{\rm AA}]=P[{\rm A}]-{\Gamma\over{\rm tr}\{{\cal X}_{\rm AB}^+\rho^{\rm ss}\}}\ ,\\
&&P[{\rm BB}]=P[{\rm B}]-{\Gamma\over{\rm tr}\{{\cal X}_{\rm AB}^+\rho^{\rm ss}\}}\ ,\\
&&P[{\rm AB}]=P[{\rm BA}]={\Gamma\over{\rm tr}\{{\cal X}_{\rm AB}^+\rho^{\rm ss}\}}\ .
\label{pab}
\end{eqnarray}
Eq.~(\ref{pab}) provides us with an interpretation of $\Gamma$ as relating to the probability that a switch occurs
in the type of detection event, given by
\begin{eqnarray}
P[{\rm AB}]+P[{\rm BA}]={2\Gamma\over{\rm tr}\{{\cal X}_{\rm AB}^+\rho^{\rm ss}\}}\ .
\label{switch}
\end{eqnarray}
Note also that these probabilities obey a useful ``distributive property''.  For example, \begin{eqnarray} P[{\rm
AA}]+P[{\rm AB}]=P[{\rm A}({\rm A}+{\rm B})]=P[{\rm A}]\ , \end{eqnarray} which is a way of stating that the detection
of an atom in state $|{\rm A}\rangle$ will necessarily be followed by the detection of an atom in either state $|{\rm
A}\rangle$ or $|{\rm B}\rangle$ with unit probability. This distributive property holds for all sequence probabilities.

Expressions for sequence probabilities for $n=3$ and greater can be formed by a sequence of the operators ${\cal
X}_{\rm A}^+$ and ${\cal X}_{\rm B}^+$ in right to left order, separated by the operator $-1/{\cal X}_{\rm AB}^-$.  For
example,
\begin{eqnarray}
P[{\rm ABB}]={1\over{\rm tr}\{{\cal X}_{\rm AB}^+\rho^{\rm ss}\}}{\rm tr}\left\{{\cal X}_{\rm B}^+{-1\over{\cal
X}_{\rm AB}^-}{\cal X}_{\rm B}^+{-1\over{\cal X}_{\rm AB}^-}{\cal X}_{\rm A}^+\rho^{\rm ss}\right\}\ ,
\end{eqnarray}
stated without proof.

\subsection{Successive detections}
\label{succsdets}

The mean number of successive detections of atoms in the same state, $\langle n\rangle$, was first calculated by Wagner
{\it et~al.}~\cite{Wagner1992,Wagner1992b,Wagner1993} using Monte Carlo techniques for the phase-sensitive micromaser
experiment described in those papers.  Later, Englert {\it et~al.}~\cite{Englert1996} derived an expression for
$\langle n\rangle$ and $\langle n\rangle_{\rm norm}$.  Using an eigenvalue method, they reproduced the results of Wagner
{\it et~al.}\ with good agreement and calculated $\langle n\rangle_{\rm norm}$ for the standard micromaser experiment. 
Here we will derive the mean number $\langle n_{\rm A}\rangle$ of successive detections of atoms in state $|{\rm
A}\rangle$ and the mean number $\langle n_{\rm B}\rangle$ of successive detections of atoms in state $|{\rm B}\rangle$.
 The derivation of $\langle n\rangle={1\over 2}(\langle n_{\rm A}\rangle+\langle n_{\rm B}\rangle)$ in
Ref.~\cite{Englert1996} needs to be modified only slightly in order to obtain $\langle n_{\rm A}\rangle$ and $\langle
n_{\rm B}\rangle$.

Given that the detection of an atom in state $|{\rm B}\rangle$ has occurred, we denote the probability of detecting $n$
atoms in state $|{\rm A}\rangle$ prior to the next detection of an atom in state $|{\rm B}\rangle$ by $p_n^{\rm
A}=P[\underline{\rm B}{\rm A}^n{\rm B}]$, where ``${\rm A}^n$'' is short for $n$ terms of ``A'' in the sequence. 
(similarly for $p_n^{\rm B}$).  They obey the normalization
\begin{eqnarray}
\sum_{n=0}^{\infty}p^{\rm A}_n=\sum_{n=0}^{\infty}p^{\rm B}_n=1\ . \label{norms}
\end{eqnarray}
Note that the probability of having n detections of atoms in one state between two detections of atoms in the other
state, $p_n$, used in the derivation of $\langle n\rangle$ in Ref.~\cite{Englert1996} is related to $p^{\rm A}_n$ and
$p^{\rm B}_n$ by \begin{eqnarray}
p_n=P[\rm B]\,p^{\rm A}_n+P[{\rm A}]\,p^{\rm B}_n=P[{\rm B}{\rm A}^n{\rm B}]+P[{\rm A}{\rm B}^n{\rm A}]\ .
\end{eqnarray}

We then consider the probability $p^{\rm A}_0=P[\underline{\rm B}{\rm B}]$ of detecting an atom in state $|{\rm
B}\rangle$ after a previous atom detected in state $|{\rm B}\rangle$, and the possible ways in which the two events can
occur between the detection of two atoms in state $|{\rm A}\rangle$:
\begin{eqnarray}
p^{\rm A}_0=P[\underline{{\rm B}}{\rm B}]&=&P[{\rm A}\underline{{\rm B}}{\rm BA}]\nonumber\\
&&{}+(P[{\rm A}\underline{{\rm B}}{\rm BBA}]+P[{\rm AB}\underline{{\rm B}}{\rm BA}])+\cdots\ .
\label{p0}
\end{eqnarray}
Noting that, for example,
\begin{eqnarray}
P[{\rm ABBA}]=P[{\rm A}]\,P[\underline{\rm A}{\rm BBA}]=P[{\rm B}]\,P[{\rm A}\underline{\rm B}{\rm BA}]\ ,
\end{eqnarray}
Eq.~(\ref{p0}) becomes
\begin{eqnarray}
p^{\rm A}_0&=&{P[{\rm A}]\over P[{\rm B}]}(P[\underline{\rm A}{\rm BBA}]+2\,P[\underline{\rm A}{\rm
BBBA}]+\cdots)\nonumber\\
&=&{P[{\rm A}]\over P[{\rm B}]}\sum_{n=2}^\infty(n-1)p^{\rm B}_n\ .
\end{eqnarray}
A similar expression is obtained for $p^{\rm B}_0$.  Together with Eqs.~(\ref{norms}), we have
\begin{eqnarray}
\sum_{n=1}^\infty np^{\rm A}_n={P[{\rm A}]\over P[{\rm B}]}p^{\rm B}_0-p^{\rm A}_0+1={P[{\rm A}]\over P[{\rm B}]}\ ,
\end{eqnarray}
and a similar expression for $p^{\rm B}_n$.

We now consider the probability $P^{\rm A}_n$ of detecting $n$ atoms in state $|{\rm A}\rangle$ in succession
(similarly for $P^{\rm B}_n$).  They are normalized as
\begin{eqnarray}
\sum_{n=1}^\infty P^{\rm A}_n=\sum_{n=1}^\infty P^{\rm B}_n=1\ .
\end{eqnarray}
$P^{\rm A}_n$ differs from $p^{\rm A}_n$ only by the exclusion of $n=0$, so that
\begin{eqnarray}
P^{\rm A}_n={p^{\rm A}_n\over1-p^{\rm A}_0}={P[{\rm B}]\over P[{\rm AB}]}p^{\rm A}_n\ .
\end{eqnarray}
For the mean number of successive detections of atoms in state $|{\rm A}\rangle$ we obtain
\begin{eqnarray}
\langle n_{\rm A}\rangle&\equiv&\sum_{n=1}^\infty nP^{\rm A}_n={P[{\rm A}]\over P[{\rm AB}]}={1\over\Gamma}{\rm
tr}\{{\cal X}_{\rm A}^+\rho^{\rm ss}\}\ ,
\end{eqnarray}
and similarly
\begin{eqnarray}
\langle n_{\rm B}\rangle={P[{\rm B}]\over P[{\rm AB}]}={1\over\Gamma}{\rm tr}\{{\cal X}_{\rm B}^+\rho^{\rm ss}\}\ .
\end{eqnarray}
The average of the two expressions gives
\begin{eqnarray}
\langle n\rangle={1\over2\,P[{\rm AB}]}={1\over2\Gamma}{\rm tr}\{{\cal X}_{\rm AB}^+\rho^{\rm ss}\}\ ,
\end{eqnarray}
which is the inverse of the switch probability (\ref{switch}) and is equivalent to Eq.~(3.24) of
Ref.~\cite{Englert1996}.  The uncorrelated values are
\begin{eqnarray}
&&\langle p^{\rm A}_0\rangle_{\rm uncor}=r_{\rm B}\int_0^\infty dt\,e^{-(r_{\rm A}+r_{\rm B})t}=P[{\rm B}]\ ,\nonumber\\
&&\langle n_{\rm A}\rangle_{\rm uncor}={1\over P[{\rm B}]}\ ,
\end{eqnarray}
and similarly for $\langle p^{\rm B}_0\rangle_{\rm uncor}$ and $\langle n_{\rm B}\rangle_{\rm uncor}$.  The average
of $\langle n_{\rm A}\rangle_{\rm uncor}$ and $\langle n_{\rm B}\rangle_{\rm uncor}$ gives Eq.~(3.10) of
Ref.~\cite{Englert1996}
\begin{eqnarray}
\langle n\rangle_{\rm uncor}={1\over2\,P[{\rm A}]\,P[{\rm B}]}\ .
\end{eqnarray}
Thus
\begin{eqnarray}
\langle n_{\rm A}\rangle_{\rm norm}&=&\langle n_{\rm B}\rangle_{\rm norm}=\langle n\rangle_{\rm norm}={P[{\rm
A}]\,P[{\rm B}]\over P[{\rm AB}]}\nonumber\\
&=&{{\rm tr}\left\{{\cal X}_{\rm A}^+\rho^{\rm ss}\right\}\,{\rm tr}\left\{{\cal X}_{\rm B}^+\rho^{\rm
ss}\right\}\over{\rm tr}\left\{{\cal X}_{\rm AB}^+\rho^{\rm ss}\right\}\,\Gamma}\ .
\label{meann}
\end{eqnarray}

\subsection{Waiting time statistics}
\label{wait}

Waiting time distributions have been calculated previously for particular trapping states~\cite{Briegel1994,Herzog1994}. 
Here we will calculate the mean waiting times between various detection events.

We begin by deriving an expression for the mean time $\langle t_{\rm A\rightarrow A}\rangle$ between detections of
atoms in state $|{\rm A}\rangle$.  Given that an atom in state $|{\rm A}\rangle$ was detected at time $0$ we have
$\rho(0)={\cal X}_{\rm A}^+\rho^{\rm ss}/{\rm tr}\{{\cal X}_{\rm A}^+\rho^{\rm ss}\}$.  At a later time $t$, the next
atom in state $|{\rm A}\rangle$ is detected.  Until then, we whish to exclude detections of atoms in state $|{\rm
A}\rangle$, but we do not care how many atoms are detected in state $|{\rm B}\rangle$.  So we use the non-normalized
conditioned density operator ${\tilde{\rho}}_{\rm c}(t)=e^{{\cal X}_{\rm A}^-rt}\rho(0)$.  Then the probability density
for the time $t$ when the next atom is detected in state $|{\rm A}\rangle$ is $r\,{\rm tr}\{{\cal X}_{\rm
A}^+{\tilde{\rho}}_{\rm c}(t)\}$.  The mean time between detections of atoms in state $|{\rm A}\rangle$ is then
\begin{eqnarray}
\langle t_{\rm A\rightarrow A}\rangle=r\int_0^\infty dt\,t\,{\rm tr}\{{\cal X}_{\rm A}^+{\tilde{\rho}}_{\rm
c}(t)\}={1\over r_{\rm A}}\,{\rm tr}\left\{{\cal X}_{\rm A}^+{1\over{{\cal X}_{\rm A}^-}^2}{\cal X}_{\rm
A}^+\rho^{\rm ss}\right\}\ .
\end{eqnarray}
Using ${\cal X}_{\rm A}^+={\cal X}-{\cal X}_{\rm A}^-$ and the identities (\ref{identities}) we have
\begin{eqnarray}
{\rm tr}\left\{{\cal X}_{\rm A}^+{1\over{{\cal X}_{\rm A}^-}^2}{\cal X}_{\rm A}^+\rho^{\rm ss}\right\}=1\ ,
\end{eqnarray}
and we conclude that the mean time between detections of atoms in state $|{\rm A}\rangle$ is equal to the uncorrelated
value
\begin{eqnarray}
\langle t_{\rm A\rightarrow A}\rangle_{\rm uncor}&=&r_{\rm A}\int_0^\infty dt\,te^{-r_{\rm A}t}={1\over r_{\rm A}}\ .
\end{eqnarray}
Likewise, for atoms in state $|{\rm B}\rangle$, $\langle t_{\rm B\rightarrow B}\rangle=\langle t_{\rm B\rightarrow
B}\rangle_{\rm uncor}=1/r_{\rm B}$.

These results do not prove that there are no correlations between atoms in the same state---only that we have to go to
higher powers in the waiting time in order to see those correlations.  For example, the mean squared time between
detections of atoms in state $|{\rm A}\rangle$ is
\begin{eqnarray}
\langle t^2_{\rm A\rightarrow A}\rangle&=&r\int_0^\infty dt\,t^2\,{\rm tr}\{{\cal X}_{\rm A}^+{\tilde{\rho}}_{\rm
c}(t)\}\nonumber\\
&=&{2\over rr_{\rm A}}{\rm tr}\left\{{\cal X}_{\rm A}^+{-1\over{{\cal X}_{\rm A}^-}^3}{\cal X}_{\rm
A}^+\rho^{\rm ss}\right\}\nonumber\\
&=&{2\over rr_{\rm A}}{\rm tr}\left\{{-1\over{\cal X}_{\rm A}^-}\rho^{\rm ss}\right\}\ ,
\end{eqnarray}
where, in the last step, we have again substituted ${\cal X}_{\rm A}^+={\cal X}-{\cal X}_{\rm A}^-$ and used the
identities (\ref{identities}).  This does not reduce to the uncorrelated value
\begin{eqnarray}
\langle t^2_{\rm A\rightarrow A}\rangle_{\rm uncor}={2\over {r_{\rm A}}^2}\ .
\end{eqnarray}

If we want a more interesting mean waiting time (one that exhibits correlations), we have to compute $\langle t_{\rm
A\rightarrow B}\rangle$ or $\langle t_{\rm B\rightarrow A}\rangle$.  Starting with the same initial condition
$\rho(0)={\cal X}_{\rm A}^+\rho^{\rm ss}/{\rm tr}\{{\cal X}_{\rm A}^+\rho^{\rm ss}\}$, and using the non-normalized
conditioned density operator ${\tilde{\rho}}_{\rm c}(t)=e^{{\cal X}_{\rm B}^-rt}\rho(0)$ for the absence of detections
of atoms in state $|{\rm B}\rangle$, the probability density for the time $t$ until the next detection of an atom in
state $|{\rm B}\rangle$ is $r\,{\rm tr}\{{\cal X}_{\rm B}^+{\tilde{\rho}}_{\rm c}(t)\}$.  The mean time until the next
detection of an atom in state $|{\rm B}\rangle$ is
\begin{eqnarray}
\langle t_{\rm A\rightarrow B}\rangle=r\int_0^\infty dt\,t\,{\rm tr}\{{\cal X}_{\rm A}^+{\tilde{\rho}}_{\rm
c}(t)\}={1\over r_{\rm A}}\,{\rm tr}\left\{{\cal X}_{\rm B}^+{1\over{{\cal X}_{\rm B}^-}^2}{\cal X}_{\rm
A}^+\rho^{\rm ss}\right\}\ .
\end{eqnarray}
Using ${\cal X}_{\rm B}^+={\cal X}-{\cal X}_{\rm B}^-$ and the identities
(\ref{identities}), we have
\begin{eqnarray}
{\rm tr}\left\{{\cal X}_{\rm B}^+{1\over{{\cal X}_{\rm B}^-}^2}{\cal X}_{\rm A}^+\rho^{\rm ss}\right\}={\rm
tr}\{{-1\over{\cal X}_{\rm B}^-}{\cal X}_{\rm A}^+\rho^{\rm ss}\}\ ,
\end{eqnarray}
so that
\begin{eqnarray}
\langle t_{\rm A\rightarrow B}\rangle={1\over r_{\rm A}}\,{\rm tr}\{{-1\over{\cal X}_{\rm B}^-}{\cal X}_{\rm
A}^+\rho^{\rm ss}\}\ ,
\end{eqnarray}
and similarly for $\langle t_{\rm B\rightarrow A}\rangle$.  The uncorrelated value is
\begin{eqnarray}
\langle t_{\rm A\rightarrow B}\rangle_{\rm uncor}=r_{\rm B}\int_0^\infty dt\,te^{-r_{\rm B}t}={1\over r_{\rm B}}\ ,
\end{eqnarray}
so that
\begin{eqnarray}
\langle t_{\rm A\rightarrow B}\rangle_{\rm norm}={r_{\rm B}\over r_{\rm A}}\,{\rm tr}\{{-1\over{\cal X}_{\rm B}^-}{\cal
X}_{\rm A}^+\rho^{\rm ss}\}\ ,
\end{eqnarray}
and similarly for $\langle t_{\rm B\rightarrow A}\rangle$.

\subsection{Numerical integration method}
\label{numerical}

Traces of the form
\begin{eqnarray}
X={\rm tr}\left\{{\cal O}_1{1\over{\cal O}_2}{\cal O}_3\rho^{\rm ss}\right\}
\end{eqnarray}
can be evaluated by first numerically integrating the linear equation (\ref{noNormConditionedMaster})
\begin{eqnarray}
\dot{\tilde{\rho}}_{\rm c}(t)=-{\cal O}_2\tilde{\rho}_{\rm c}(t)\ ,\ \ \tilde{\rho}_{\rm c}(0) =
{\cal O}_3\rho^{\rm ss}\ ,
\end{eqnarray}
to obtain $\tilde{\rho}_{\rm c}(t)$, and then integrating the equation
\begin{eqnarray}
\dot{X}(t)={\rm tr}\{{\cal O}_1\tilde{\rho}_{\rm c}(t)\}\ ,\ \ X(0)=0\ ,
\end{eqnarray}
to obtain $X=\lim_{t\rightarrow\infty}X(t)$.

When the eigenvalues and eigenstates of the operator ${\cal O}_2$ are known analytically, the eigenvalue method used
in Ref.~\cite{Englert1996} is much faster.  However, for the general case, the eigenvalues and eigenstates must be
obtained numerically.  If one does not care about the eigenvalues, then numerical integration is simpler to implement
and likely to be faster.  In calculating the mean number of successive detections of atoms in the same state, the
Authors of Ref.~\cite{Englert1996} state that numerical integration is faster ``roughly by a factor of  two'', but
neither the numerical integration method nor the method for obtaining the eigenvalues and eigenstates were clearly
stated.

As an alternative to numerical methods, McGowan and Schieve~\cite{McGowan1999} have used a perturbation method to
derive approximate analytical solutions for $\langle n\rangle$.  This is a general method that could be used, in
principle, to calculate other statistical quantities, though it is unclear as to when the method is a valid
approximation.

\section{Standard Micromaser Experiment}
\label{standSetup}

Excluding added complications (e.g. two atom events or non-Poissonian injection statistics), the methods for deriving
various statistical quantities describing the detection of the emerging atoms (Sec.~\ref{detStats}) are completely
general, as are the expressions given in Sec.~\ref{observables}.  For a particular micromaser experiment only the
super operators ${\cal A}$ and ${\cal B}$, and the steady state density operator $\rho^{\rm ss}$ need to be specified. 
Here we will give results only for the standard micromaser arrangement in which the atoms arrive in the upper maser
level, and the excitation level of the emerging atoms are measured.  Other micromaser experiments may be considered
in future papers.

We have allready specified $\rho^{\rm ss}$ (Eq. (\ref{steady})).  All that remains is to specify ${\cal A}$ and ${\cal
B}$.  Identifying the states $|{\rm A}\rangle$ and $|{\rm B}\rangle$ with the upper and lower maser levels,
respectively, ${\cal A}$ and ${\cal B}$ are given by~\cite{Scully1997}
\begin{eqnarray}
{\cal
A}\rho&=&\cos(\varphi\sqrt{aa^\dagger})\rho\cos(\varphi\sqrt{aa^\dagger})\ ,\nonumber\\ {\cal
B}\rho&=&\hat{\gamma}^\dagger\sin(\varphi\sqrt{aa^\dagger})\rho\sin(\varphi\sqrt{aa^\dagger})\hat{\gamma}\ .
\end{eqnarray}
Note that only the diagonal elements of the operators ${\cal A}$, ${\cal B}$, and ${\cal L}$ are needed,
since $\rho^{\rm ss}$ is diagonal, and these operators preserve diagonality.

All numerical results presented in this paper assumed symmetric detector efficiencies $\eta_{\rm A}=\eta_{\rm
B}=\eta$, though the methods can be used for non-symmetric detector efficiencies.

\subsection{Atomic inversion}

Trapping states were recently observed by Weidinger {\it et~al.}~\cite{Weidinger1999} in measurements of the atomic
inversion.  In that paper, the atomic inversion is defined as the difference in the probabilities of a given
injected atom being detected in the ground state and the excited state
\begin{eqnarray}
I\equiv{\rm tr}\{({\cal X}_{\rm B}^+-{\cal X}_{\rm A}^+)\rho^{\rm ss}\}={1\over r}(r_{\rm B}-r_{\rm A})\ .
\end{eqnarray}
Another definition is the difference in the probabilities of a given {\em detected} atom being in the ground state and
the excited state
\begin{eqnarray}
\tilde{I}\equiv P[{\rm B}]-P[{\rm A}]={r_{\rm B}-r_{\rm A}\over r_{\rm A}+r_{\rm B}}\ .
\end{eqnarray}
Both formulas reduce to the difference in the number of atoms in the ground state and excited state ${\rm tr}\{({\cal
B}-{\cal A})\rho^{\rm ss}\}$ in the ideal limit of one hundred percent detector efficiencies.

\begin{figure}
\begin{picture}(200,310)(0,0)
\put(110,150){\makebox(0,0){\scalebox{0.43}{\includegraphics*[15mm,-10mm][200mm,260mm]{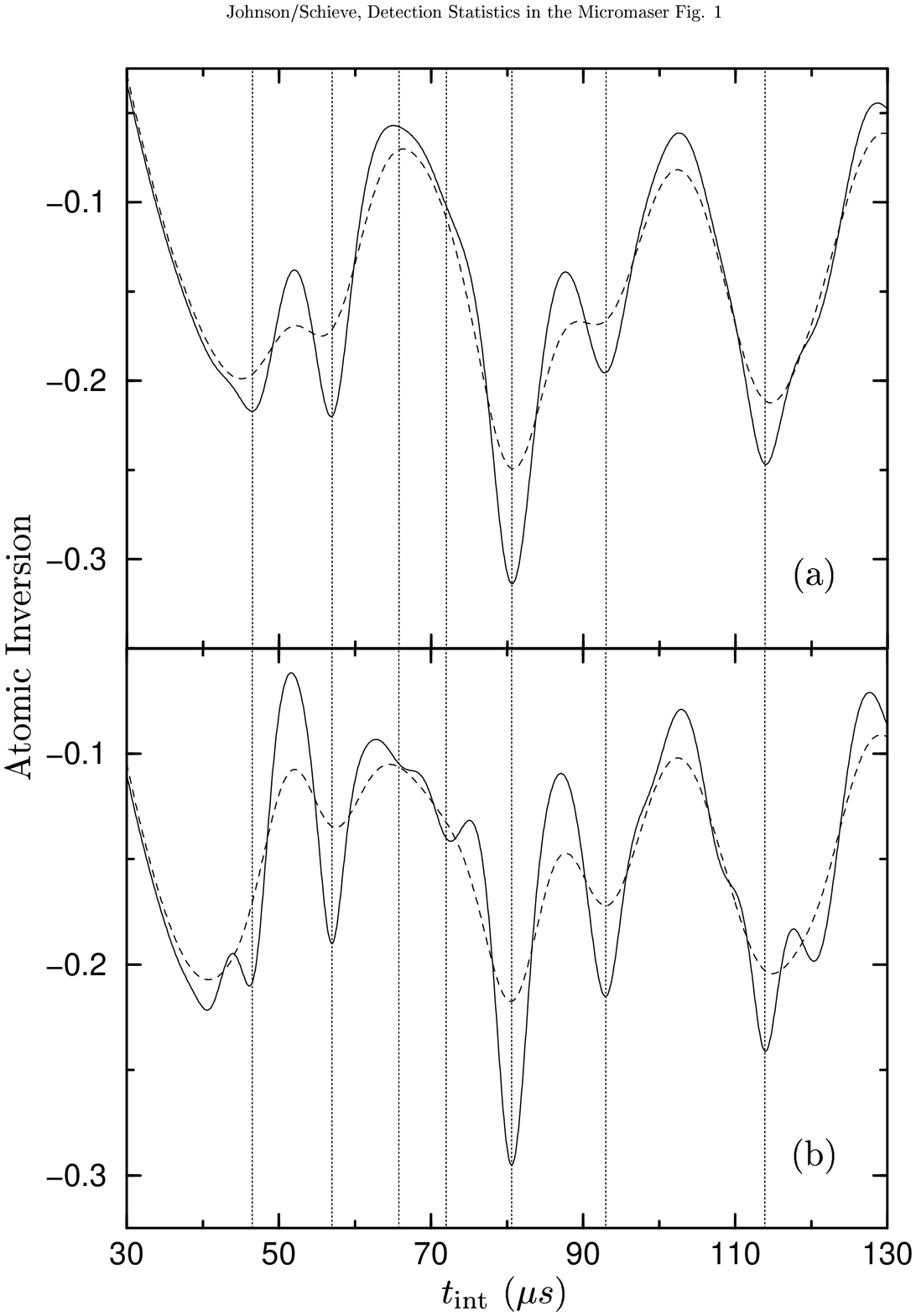}}}}
\end{picture}
\caption{Plots (a) and (b) show the atomic inversion for $N_{\rm ex}=7$ and $N_{\rm ex}=10$ respectively, for comparison
to Figs.~3($\alpha$) and~3($\beta$) of Ref.~\protect\cite{Weidinger1999}.  Here $\nu=0.054$ and $\eta=40\%$.  The
vertical dotted lines indicate the same trapping states as in Fig.~3 of Ref.~\protect\cite{Weidinger1999}.  The dashed
lines show the effect of a $3\ \mu s$ uncertainty in $t_{\rm int}$ with gaussian distribution.}
\label{inversion}
\end{figure}

Fig.~\ref{inversion} shows $I$ calculated for $N_{\rm ex}=7$ and $N_{\rm ex}=10$ for comparison to Figs.~3($\alpha$)
and~3($\beta$) of Ref.~\cite{Weidinger1999}.  Here $\nu=0.054$ and $\eta=40\%$.  The coupling constant for the
$63P_{3/2}-61D_{5/2}$ Rydberg transition in rubidium was determined in Ref.~\cite{Weidinger1999} as $g=39\ \pm\ 5$ kHz.
 Here we have used $g=39$ kHz to plot $I$ as a function of $t_{\rm int}$.  The vertical dotted lines indicate the same
trapping states as in Fig.~3 of Ref.~\cite{Weidinger1999}.  It should be noted that the linear trend present in the raw
experimental data due to the unequal spontaneous emission rates and its subsequent removal makes the scale on the
vertical axis on Figs. 3($\alpha$) and 3($\beta$) of Ref.~\cite{Weidinger1999} useful only for comparisons of relative
heights.  There is qualitative agreement between the theoretical and experimental results in that each set of plots has
dips centered on the trapping states and approximately the same amplitude of modulation, but little quantitative
agreement.  As a simple attempt to improve the agreement, we added the effect of an uncertainty to $t_{\rm int}$ with a
gaussian distribution to simulate velocity averaging in the experiment.  The dashed curves on Fig.~\ref{inversion} are
the effect of an uncertainty of $3\ \mu s$.  This has the effect of producing a level of amplitude suppression
comparable to that for the experimental data.  Other effects that we added (but do not include the results here) were
the spontaneous emission of the atoms and the removal of the subsequent linear trend, and a misalignment of the
detection unit (which would cause incorrect detection) as discussed in Sec.~3.1.4 of Ref.~\cite{Englert1998}.  The
inclusion of either of these effects did not cause any significant change in the shape of the graphs.  A more involved
parameterized fitting of the experimental results correcting for the effects of detuning, spatial dependence of the
atom-field coupling, and others (see the discussion of Eq.~(73) in Ref.~\cite{Englert1998}) may prove useful, but will
not be attempted here.  It should be emphasized that this lack of quantitative agreement is within the margins of
statistical error of the measurements, and does not negate the observation of trapping states or indicate any problems
with the theory.

\subsection{Fano-Mandel functions}

\begin{figure}
\begin{picture}(200,175)(0,0)
\put(105,20){\makebox(0,0){\scalebox{0.43}{\includegraphics*[15mm,-10mm][200mm,260mm]{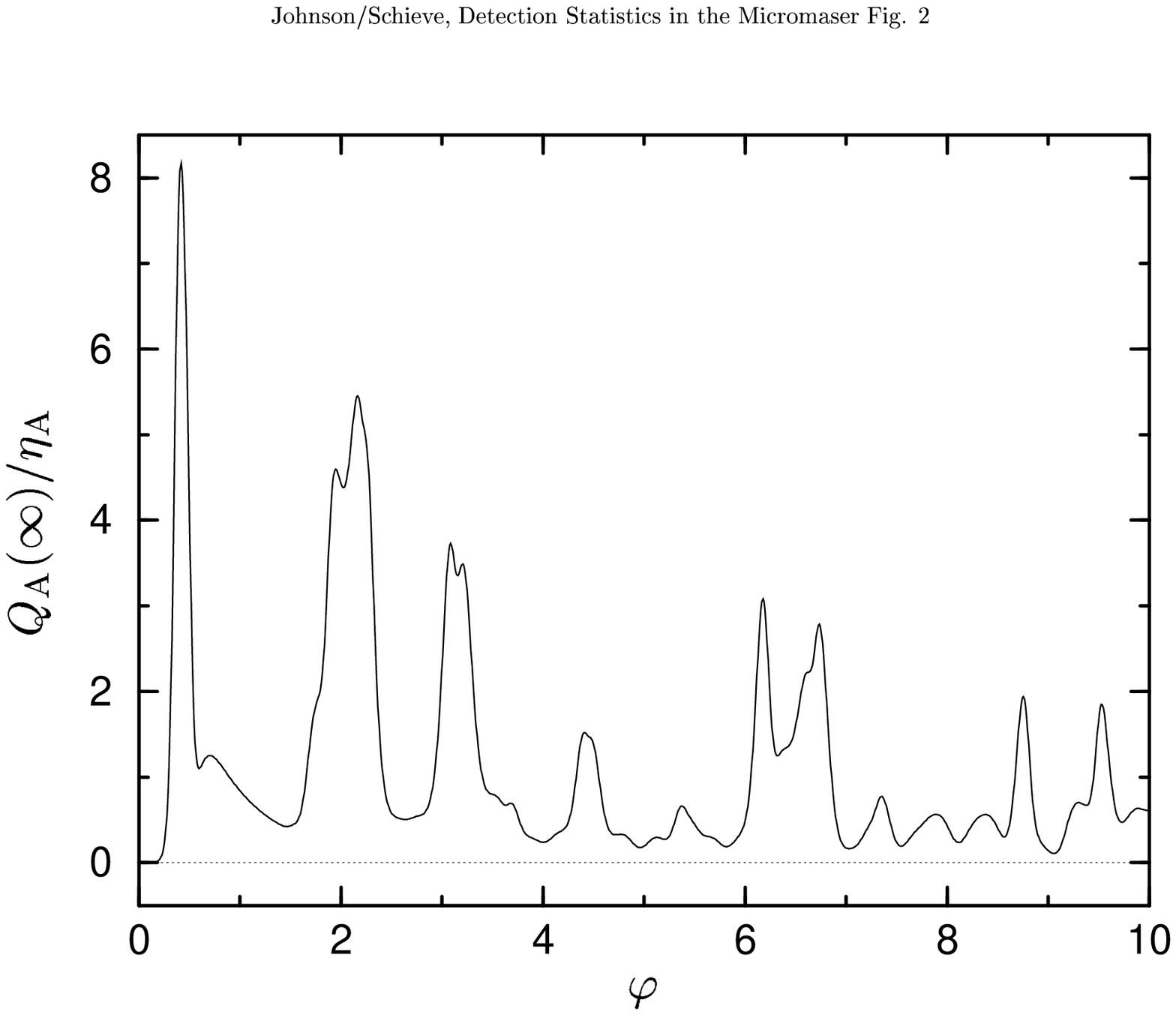}}}}
\end{picture}
\caption{Steady state Fano-Mandel function $Q_{\rm A}(\infty)$ for detections of excited atoms for $N_{\rm ex}=7$ and
$\nu=0.054$.  The result is independent of the detector efficiencies when scaled by $1/\eta_{\rm A}$.}
\label{qa}
\end{figure}

\begin{figure}
\begin{picture}(200,175)(0,0)
\put(105,20){\makebox(0,0){\scalebox{0.43}{\includegraphics*[15mm,-10mm][200mm,260mm]{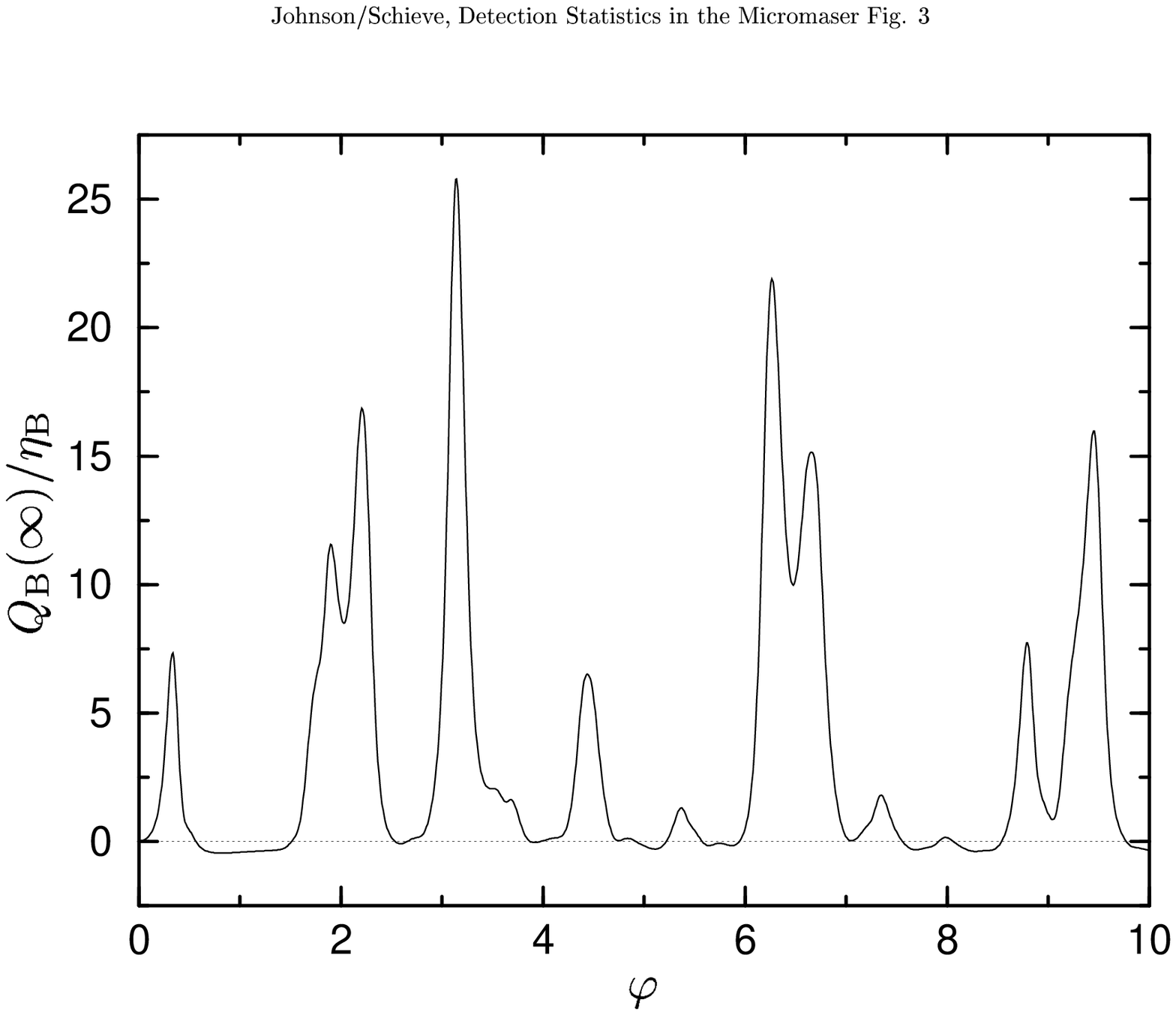}}}}
\end{picture}
\caption{Steady-state Fano-Mandel function $Q_{\rm B}(\infty)$ for detections of de-excited atoms for $N_{\rm ex}=7$ and
$\nu=0.054$.  The result is independent of the detector efficiencies when scaled by $1/\eta_{\rm B}$.}
\label{qb}
\end{figure}

Figs.~\ref{qa} and \ref{qb} show the steady state Fano-Mandel functions $Q_{\rm A}(\infty)$ and $Q_{\rm B}(\infty)$,
scaled by $1/\eta_{\rm A}$ and $1/\eta_{\rm B}$ respectively.  The results are independent of detector efficiencies
when scaled in this way.  The parameters used here and in all subsequent plots are $N_{\rm ex}=7$ and $\nu=0.054$. 
Fig.~\ref{qb} shows many regions of sub-Poissonian fluctuations in the detection of de-excited atoms.  $\nu=0.054$
is enough to destroy the visibility of most trapping states (not shown).  For smaller values of $\nu$, trapping states
have the effect of reducing the fluctuations of counts, sometimes to sub-Poissonian levels, creating dips in the
Fano-Mandel functions.  Exceptions occur for $Q_{\rm B}(\infty)$ at the trapping states $\varphi=\pi,\ 2\pi,\
3\pi,\ldots,$, where maxima occur in the fluctuations.

\begin{figure}
\begin{picture}(200,175)(0,0)
\put(105,20){\makebox(0,0){\scalebox{0.43}{\includegraphics*[15mm,-10mm][200mm,260mm]{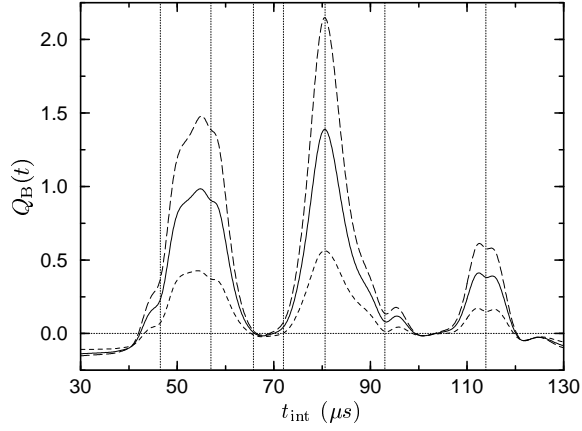}}}}
\end{picture}
\caption{Fano-Mandel function $Q_{\rm B}(t)$ for detections of de-excited atoms for comparison to Fig.~4 of
Ref.~\protect\cite{Weidinger1999}.  Here $N_{\rm ex}=7$ and $\nu=0.054$.  The long-dashed line shows $Q_{\rm
B}(t=4/\gamma)$, the dashed line shows $Q_{\rm B}(t=1/\gamma)$, and the solid line shows the result of averaging
$Q_{\rm B}(t)$ for times between one and four cavity decay times.  The vertical dotted lines indicate the same trapping
states as in Fig.~4 of Ref.~\protect\cite{Weidinger1999}.}
\label{qbCompare}
\end{figure}

Fig.~\ref{qbCompare} shows $Q_{\rm B}(t)$ for observation times of one and four cavity decay times, as well as an
average of $Q_{\rm B}(t)$ for observation times between one and four cavity decay times.  This is to be compared to
Fig.~4 of Ref.~\cite{Weidinger1999}, where the points plotted were averaged over time periods between one and four
cavity decay times.  Here, we again have $\eta=40\%$ and have used $g=39$ kHz to plot $Q_{\rm B}(t)$ as a function of
$t_{\rm int}$.  The vertical dotted lines indicate the same trapping states as in Fig.~4 of Ref.~\cite{Weidinger1999}. 
In sub-Poissonian regions, the Fano-Mandel function reaches a steady state value at about one cavity decay
time~\cite{Briegel1994}, which can be seen here by the small amount of differences between $Q_{\rm B}(t=1/\gamma)$ and
$Q_{\rm B}(t=4/\gamma)$ in the sub-Poissonian regions.  In the super-Poissonian regions, $Q_{\rm B}(t)$ approaches its
steady state value at a much slower rate.

Compared to the theoretical calculation, the experimental results exhibit an amplitude suppression of roughly a factor
of ten.  Sources of amplitude suppression include velocity averaging and the spontaneous decay of the atoms during the
times between leaving the cavity and reaching the detector.  A discussion of these added effects and the resulting
fluctuations may be included in a later paper, but will not be included here.  Without the inclusion of these effects,
the comparison may not be a meaningful indicator of the amount of agreement between the experimental results and the
theory.

\subsection{Successive detections}

\begin{figure}
\begin{picture}(200,175)(0,0)
\put(105,20){\makebox(0,0){\scalebox{0.43}{\includegraphics*[15mm,-10mm][200mm,260mm]{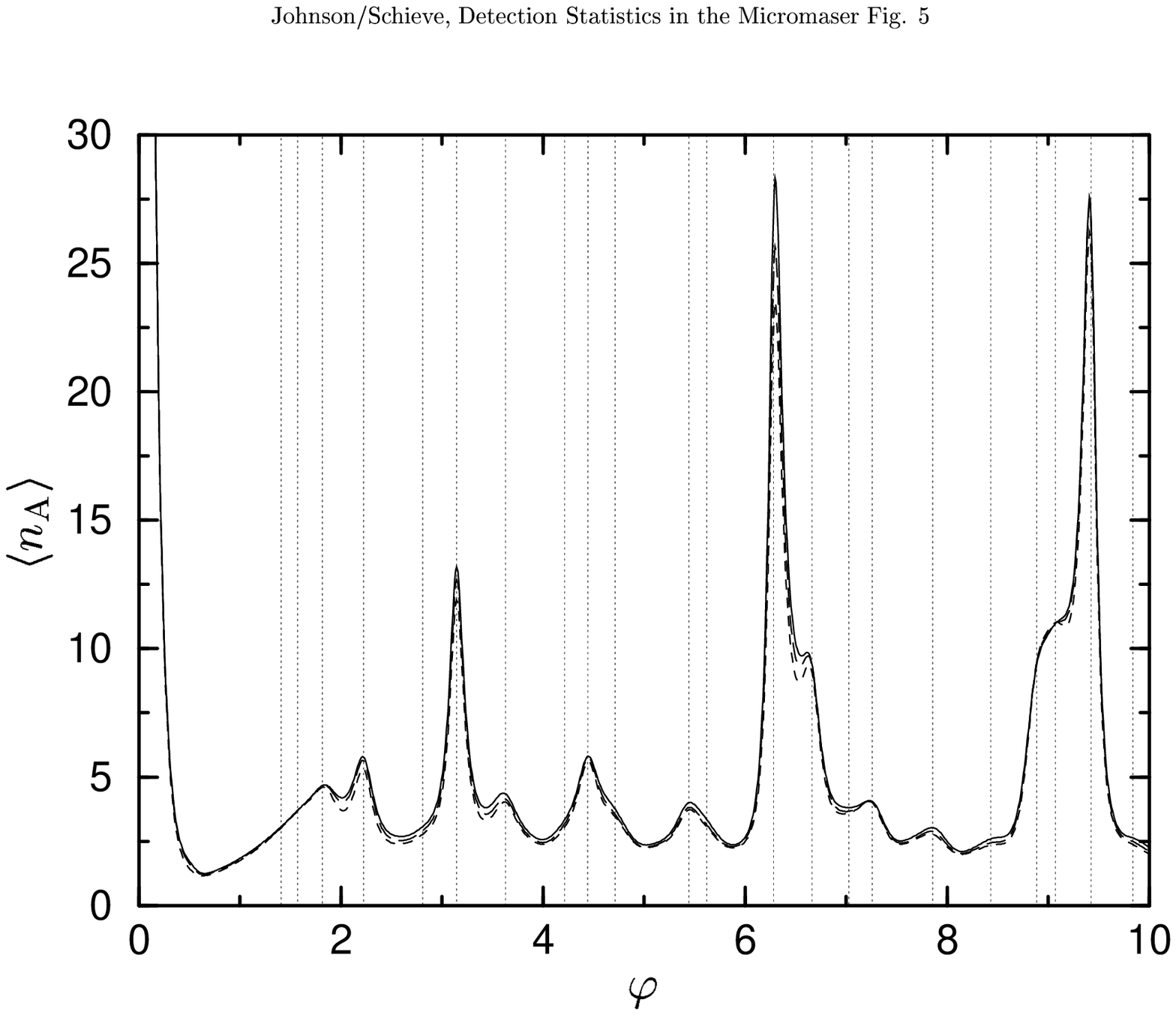}}}}
\end{picture}
\caption{Mean number $\langle n_{\rm A}\rangle$ of successive detections of excited atoms for $N_{\em ex}=7$ and
$\nu=0.054$.  The curves are for $\eta=100\%$ (solid line), $\eta=40\%$ (long dashed line), and $\eta=10\%$ (dashed
line).  The vertical dotted lines indicate the trapping states for $n_0=0,\ldots, 4$.}
\label{na}
\end{figure}

\begin{figure}
\begin{picture}(200,175)(0,0)
\put(105,20){\makebox(0,0){\scalebox{0.43}{\includegraphics*[15mm,-10mm][200mm,260mm]{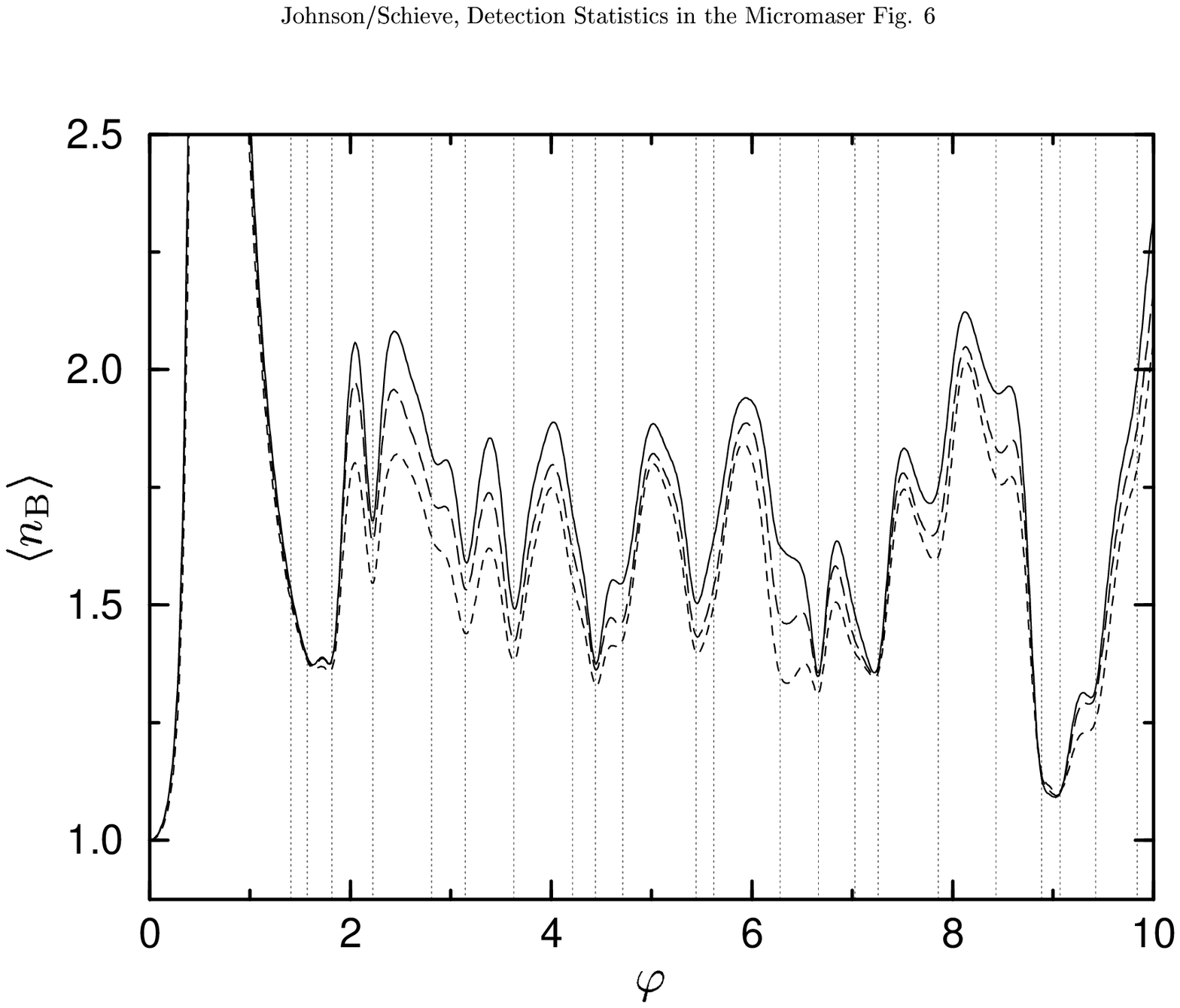}}}}
\end{picture}
\caption{Mean number $\langle n_{\rm B}\rangle$ of successive detections of de-excited atoms for $N_{\rm ex}=7$ and
$\nu=0.054$.  The curves are for $\eta=100\%$ (solid line), $\eta=40\%$ (long dashed line), and $\eta=10\%$ (dashed
line).  The vertical dotted lines indicate the trapping states for $n_0=0,\ldots, 4$.}
\label{nb}
\end{figure}

\begin{figure}
\begin{picture}(200,175)(0,0)
\put(105,20){\makebox(0,0){\scalebox{0.43}{\includegraphics*[15mm,-10mm][200mm,260mm]{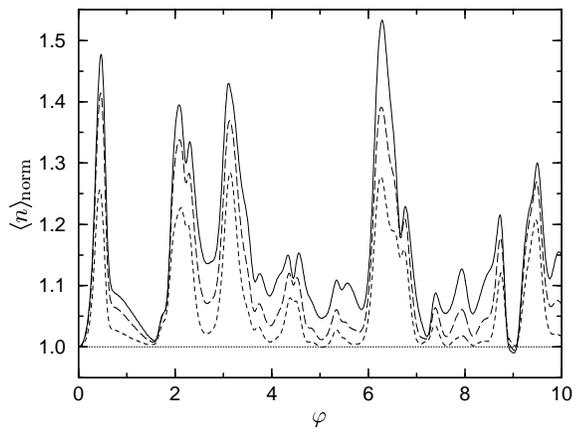}}}}
\end{picture}
\caption{Mean number of successive detections (of any type) normalized to the uncorrelated value for $N_{\rm ex}=7$ and
$\nu=0.054$.  The curves are for $\eta=100\%$ (solid line), $\eta=40\%$ (long dashed line), and $\eta=10\%$ (dashed
line).}
\label{nnorm}
\end{figure}

Fig.~\ref{na} shows the mean number of successive detections of excited atoms, $\langle n_{\rm A}\rangle$, for
$\eta=100\%$, $40\%$, and  $10\%$.  The vertical dotted lines indicate the trapping states for $n_0=0,\ldots, 4$.  At
a trapping state, if it were not for the effect of thermal damping, we would expect all of the atoms to emerge in the
upper maser level, having undergone an integer multiple of Rabi cycles.  Some trapping states give rise to large peaks
indicating an increase in the number of successive detections of excited atoms as expected.  Others have a smaller
effect on the shape of the graph that can only be seen in the change of the slope of a nearby peak.  This suggests that
the curve could be approximated by a superposition of suitably weighted gaussians centered on the trapping states. 
Fig.~\ref{nb} shows the mean number of successive detections of de-excited atoms, $\langle n_{\rm B}\rangle$.  This
graph shows the analogous effect of trapping states with the trapping states giving rise to dips rather than peaks. 
Both Figs.~\ref{na} and~\ref{nb} show a decrease in the mean number of successive detections as the detectors become
less efficient.  The uncorrelated values are unaffected by the detector efficiencies (for equal detector efficiencies),
and this decrease in the mean number of successive detections is just an indication of the decrease in correlations for
decreasing detector efficiencies.  Fig.~\ref{nnorm} shows the mean number of successive detections (of any type)
normalized to the uncorrelated value, $\langle n_{\rm A}\rangle_{\rm norm}=\langle n_{\rm B}\rangle_{\rm norm}=\langle
n\rangle_{\rm norm}$.  For decreasing detector efficiency, $\langle n\rangle$ approaches unity as expected. 
Antibunching occurs in a narrow interval of $\varphi=9$.

\subsection{Waiting times}

\begin{figure}
\begin{picture}(200,175)(0,0)
\put(105,20){\makebox(0,0){\scalebox{0.43}{\includegraphics*[15mm,-10mm][200mm,260mm]{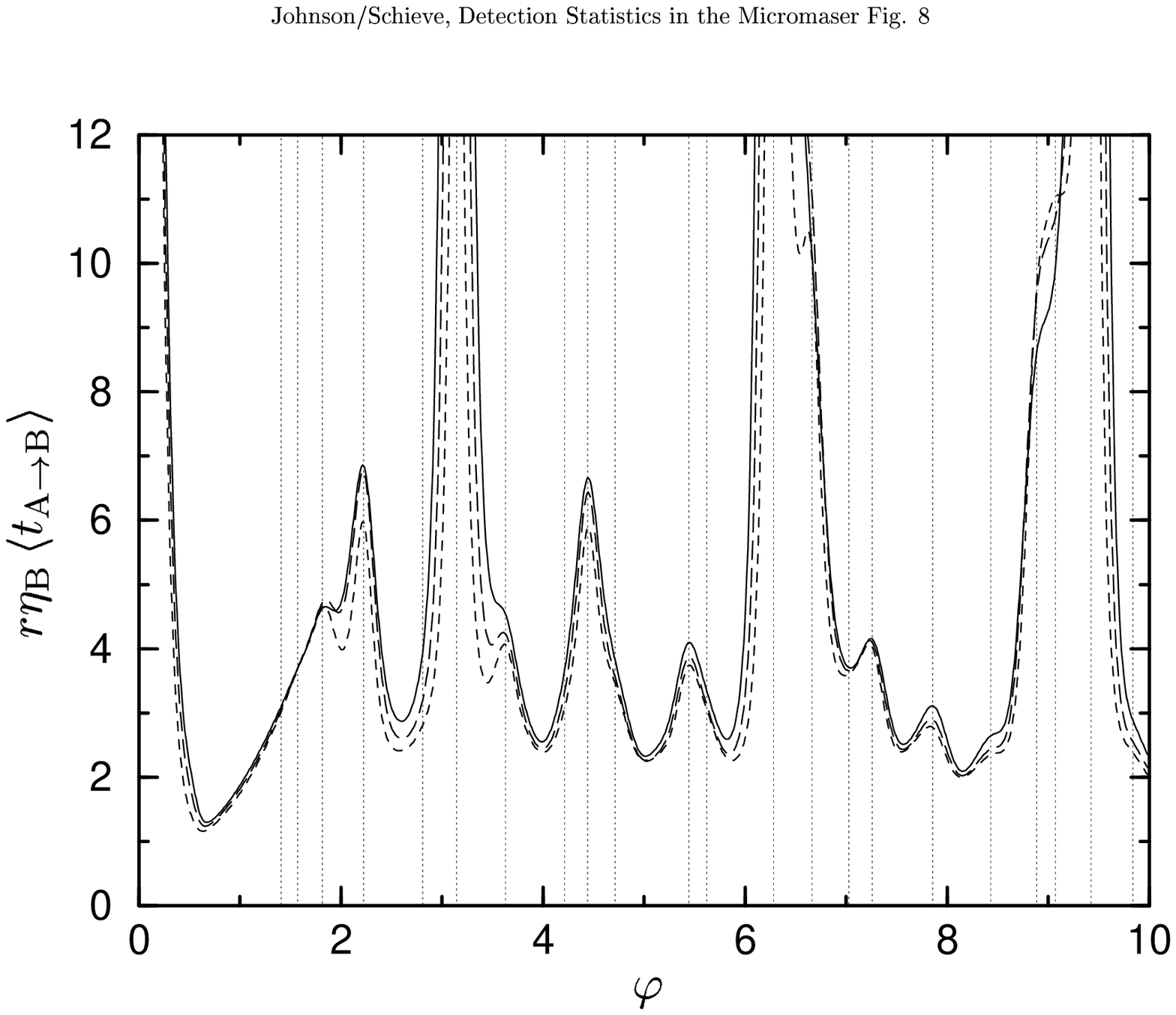}}}}
\end{picture}
\caption{Mean waiting time $\langle t_{\rm A\rightarrow B}\rangle$ until the next detection of a de-excited atom after
an initial detection of an excited atom (scaled by the $r\eta_{\rm B}$) for $N_{\rm ex}=7$ and $\nu=0.054$.  The
curves are for $\eta=100\%$ (solid line), $\eta=40\%$ (long dashed line), and $\eta=10\%$ (dashed line).  The vertical
dotted lines indicate the trapping states for $n_0=0,\ldots, 4$.}
\label{tab}
\end{figure}

\begin{figure}
\begin{picture}(200,175)(0,0)
\put(105,20){\makebox(0,0){\scalebox{0.43}{\includegraphics*[15mm,-10mm][200mm,260mm]{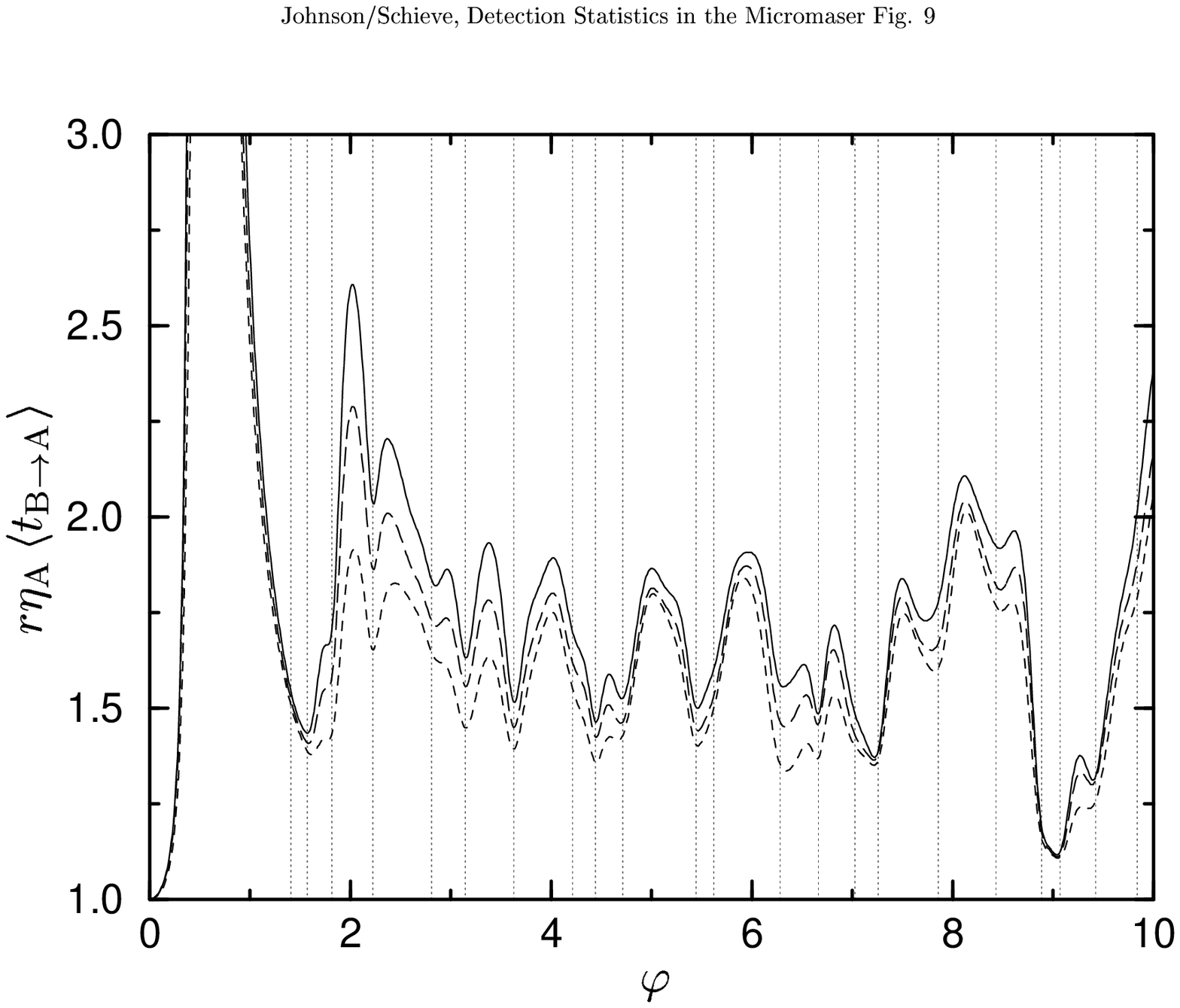}}}}
\end{picture}
\caption{Mean waiting time $\langle t_{\rm B\rightarrow A}\rangle$ until the next detection of an excited atom after an
initial detection of a de-excited atom (scaled by $r\eta_{\rm A}$) for $N_{\rm ex}=7$ and $\nu=0.054$.  The curves are
for $\eta=100\%$ (solid line), $\eta=40\%$ (long dashed line), and $\eta=10\%$ (dashed line).  The vertical dotted lines
indicate the trapping states for $n_0=0,\ldots, 4$.} \label{tba}
\end{figure}

\begin{figure}
\begin{picture}(200,175)(0,0)
\put(105,20){\makebox(0,0){\scalebox{0.43}{\includegraphics*[15mm,-10mm][200mm,260mm]{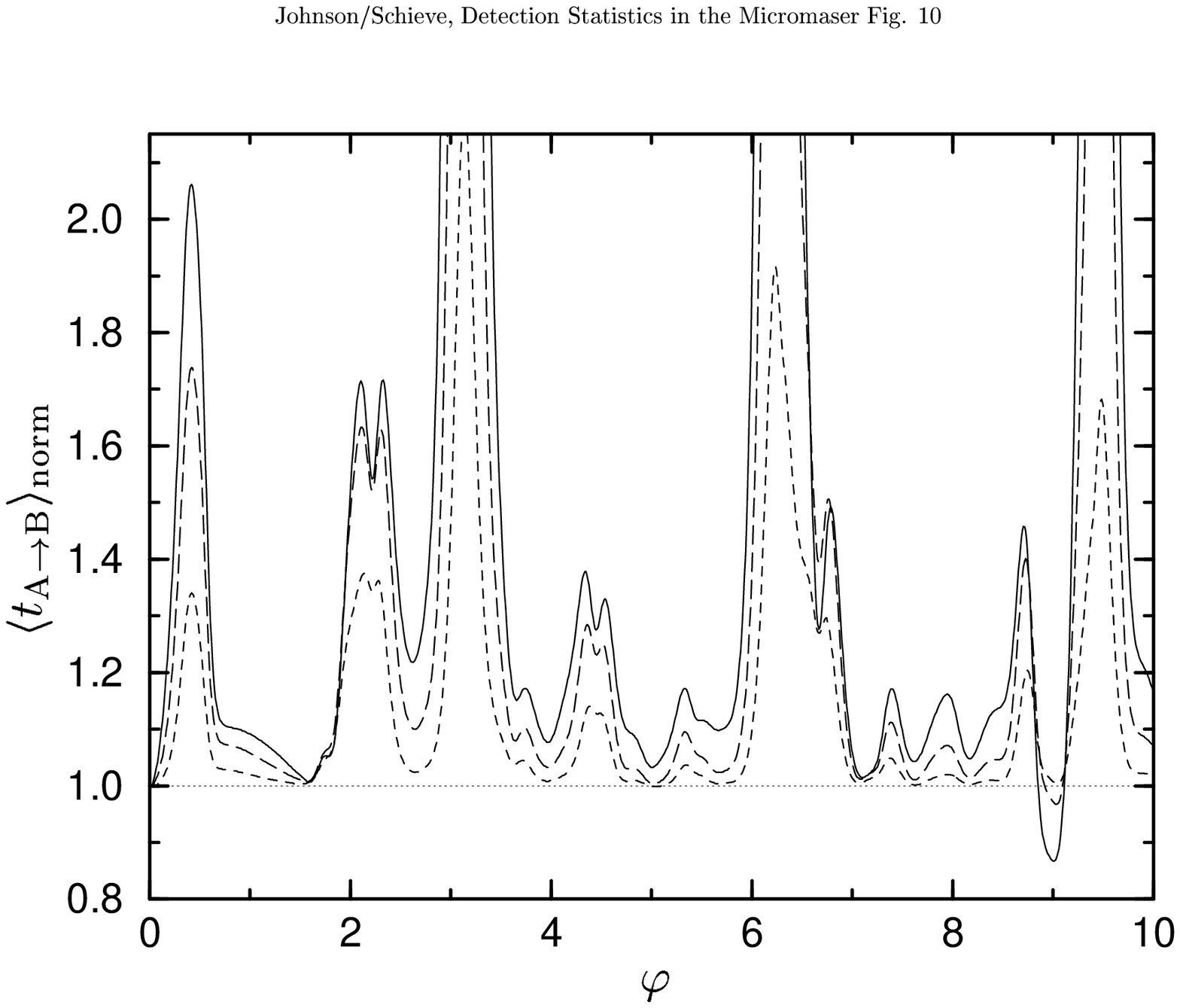}}}}
\end{picture}
\caption{Mean waiting time $\langle t_{\rm A\rightarrow B}\rangle$ until the next detection of a de-excited atom after
an initial detection of an excited atom normalized to the uncorrelated value for $N_{\rm ex}=7$ and $\nu=0.054$.  The
curves are for $\eta=100\%$ (solid line), $\eta=40\%$ (long dashed line), and $\eta=10\%$ (dashed line).}
\label{tabnorm} \end{figure}

\begin{figure}
\begin{picture}(200,175)(0,0)
\put(105,20){\makebox(0,0){\scalebox{0.43}{\includegraphics*[15mm,-10mm][200mm,260mm]{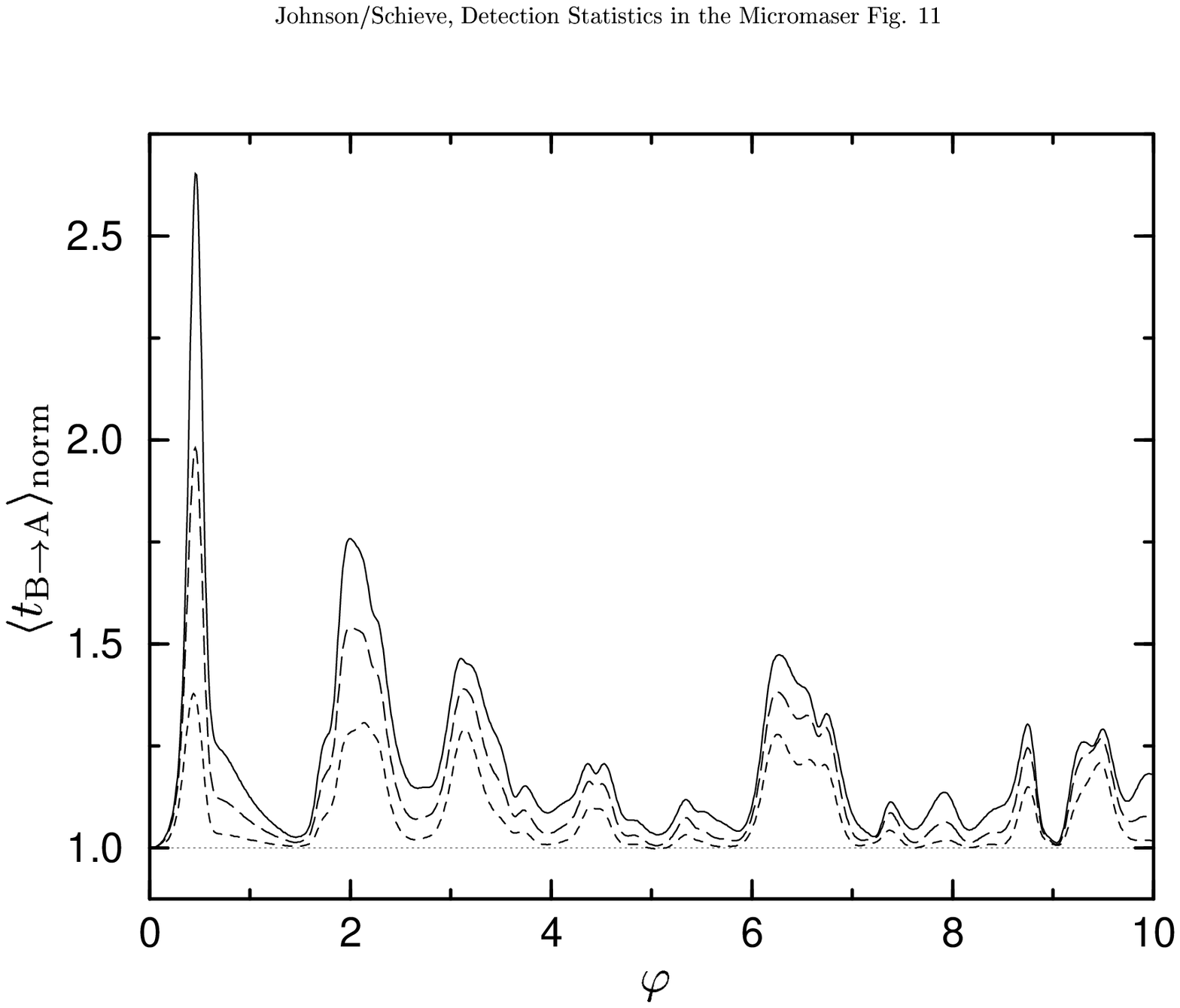}}}}
\end{picture}
\caption{Mean waiting time $\langle t_{\rm B\rightarrow A}\rangle$ until the next detection of an excited atom after an
initial detection of a de-excited atom normalized to the uncorrelated value for $N_{\rm ex}=7$ and $\nu=0.054$.  The
curves are for $\eta=100\%$ (solid line), $\eta=40\%$ (long dashed line), and $\eta=10\%$ (dashed line).}
\label{tbanorm}
\end{figure}

Fig.~\ref{tab} shows the average time until the next detection of a de-excited atom after an initial detection of an
excited atom, $\langle t_{\rm A\rightarrow B}\rangle$, scaled by $r\eta_{\rm B}$. The scaling compensates for the
increase in mean waiting time as the detectors become elss efficient and the probability of the final detection
occurring decreases.  Again, the vertical dotted lines indicate the trapping states for $n_0=0,\ldots, 4$.  The
trapping states have the effect of giving rise to peaks in the average waiting time for the next detection of a
de-excited atom as expected.  As before, some of the trapping states have a small effect.  Fig.~\ref{tba} shows
$r\eta_{\rm A}\langle t_{\rm B\rightarrow A}\rangle$ which shows the analogous effect of trapping states giving rise to
dips rather than peaks.  Fig.~\ref{tabnorm} and \ref{tbanorm} show $\langle t_{\rm A\rightarrow B}\rangle_{\rm norm}$
and $\langle t_{\rm B\rightarrow A}\rangle_{\rm norm}$ respectively.  Fig.~\ref{tabnorm} shows some negative
correlations in a limited vicinity of $\varphi=9$.

\section{Conclusions}

The effect of trapping states on the mean number of successive detections of atoms in a particular state and the mean
waiting times between detections of atoms in unlike states offer an alternative means of the experimental observation
of trapping states, though a comparison of the errors in calculating these quantities from experimental data is
required to determine which is most useful.  In the near future, we plan to consider the phase-sensitive setup
of~\cite{Wagner1992,Wagner1992b,Wagner1993}, in which case only the form of the operators ${\cal A}$ and ${\cal B}$
need to be altered for the expressions derived in this paper to carry over.  We may consider other experimental setups
as well.

It would appear that the theory of atomic detection statistics in the micromaser is mature, though it would be
interesting to include effects absent in the treatment of the ideal micromaser, and thus improve the chances for
realizing connections between the detection statistics and the field variables in a working micromaser.  Possibilities
include two-atom events, spontaneous emission, velocity averaging, and suppression of perfect inversion by thermal
photons, among others.  Also important is knowing which of these effects is the most significant.  We plan to
pursue this direction of research in the future.


\end{document}